# Wide-Field InfraRed Survey Telescope (WFIRST) Mission and Synergies with LISA and LIGO-Virgo

**N Gehrels[1], D Spergel[2], on behalf of the WFIRST SDT and Project**
[1] NASA Goddard Space Flight Center, Greenbelt, Maryland 20771 USA
[2] Princeton University, Peyton Hall, Princeton, NJ, 08544 USA

[1] E-mail: neil.gehrels@nasa.gov

**Abstract.** The Wide-Field InfraRed Survey Telescope (WFIRST) is a NASA space mission in study for launch in 2024. It has a 2.4 m telescope, wide-field IR instrument operating in the 0.7 - 2.0 micron range and an exoplanet imaging coronagraph instrument operating in the 400 - 1000 nm range. The observatory will perform galaxy surveys over thousands of square degrees to J=27 AB for dark energy weak lensing and baryon acoustic oscillation measurements and will monitor a few square degrees for dark energy SN Ia studies. It will perform microlensing observations of the galactic bulge for an exoplanet census and direct imaging observations of nearby exoplanets with a pathfinder coronagraph. The mission will have a robust and well-funded guest observer program for 25% of the observing time. WFIRST will be a powerful tool for time domain astronomy and for coordinated observations with gravitational wave experiments. Gravitational wave events produced by mergers of nearby binary neutron stars (LIGO-Virgo) or extragalactic supermassive black hole binaries (LISA) will produce electromagnetic radiation that WFIRST can observe.

## 1. Introduction

WFIRST is the top-ranked large space mission of the 2010 NWNH Decadal Survey. The mission is designed to settle essential questions in dark energy, exoplanets, and infrared astrophysics. The mission will feature strategic key science programs plus a vigorous program of guest observations. The observatory includes a wide-field near infrared camera to perform surveys with HST-style imaging and sensitivity, but further in the IR and with hundreds of times the sky coverage. It also will have a pathfinder coronagraph instrument to perform sensitive direct imaging of exoplanets. A Science Definition Team and a Study Office at the Goddard Space Flight Center and Jet Propulsion Laboratory are studying the mission with reports [1, 2] in 2013 and 2014.

The Decadal Survey identified a tripod of focus science themes for WFIRST in the areas of cosmology, planetary systems and general astrophysics. The cosmology theme will highlight measurements of the expansion history of the universe and growth of structure to learn about dark energy. The observations will include near infrared galaxy imaging and slitless spectroscopic surveys over 2400 $\deg^2$ in the baseline 6 year mission. The expansion history will also be studied by Type Ia supernova monitoring.

The planetary systems theme will highlight observations of exoplanets using the microlensing and coronagraphic techniques. Microlensing will be done by monitoring a 3 $\deg^2$ field in the galactic bulge in the infrared to detect microlensing magnification events caused by the alignment of a background star and foreground star with a planetary system. The events will measure the masses and distances to the planets and their host star, and provide a census of the prevalence of different kinds of exoplanets. The direct imaging observations will be performed with a coronagraph, observing known nearby exoplanets and debris disks and searching for new ones. It will be sensitive to larger planets in the Neptune-Jupiter size class.

The general astrophysics theme will highlight survey and monitoring science of sources throughout the universe. The research will be done using archival data from the dark energy and exoplanet surveys, as well as new observations performed as part of a Guest Observer program. Science examples are solar system studies of asteroids, galactic studies of stellar populations and brown dwarfs, surveys to study galaxy evolution and star formation history of the universe, and cosmological studies of the earliest stars, black holes and galaxies.

The WFIRST observatory will have a target-of-opportunity capability that will make a valuable resource for follow-up of gravitational wave events detected by LIGO-Virgo and LISA. The combination of wide field of view, high sensitivity and precise imaging will allow deep searches and observations of electromagnetic counterparts.

The observatory is described in Section 2, observing program in Section 3, science in Section 4, gravitational wave interest in Section 5 and summary in Section 6.

## 2. WFIRST Observatory

WFIRST has recently been baselined with an existing 2.4 m telescope NASA acquired from the National Reconnaissance Office, compared with a 1.5 m telescope considered in NWNH. This configuration is referred to as WFIRST-AFTA. The 2.4 m obscured telescope feeds two different instrument volumes containing the wide-field instrument and the coronagraph. The wide-field instrument includes two channels, a wide-field channel and an integral field unit (IFU) spectrograph channel. The wide-field channel includes three mirrors (two folds and a tertiary) and a filter/grism wheel to provide an imaging mode covering 0.76 – 2.0 micron and a spectroscopy mode covering 1.35 – 1.95 micron. The wide-field focal plane uses 2.5 micron long-wavelength cutoff 4k x 4k HgCdTe detectors. The HgCdTe detectors are arranged in a 6x3 array, providing an active area of 0.281 $deg^2$. The pixel scale is 0.11 arcsec and filter bandpass in microns are z(0.76-0.98), Y(0.93-1.19), J(1.13-1.45), H(1.38-1.77), Ks(1.68-2.0) and Wide (0.93-2.00). The grism has a spectral resolving power of 461*lambda (1.35 microns < lambda < 1.95 microns). The IFU channel uses an image slicer and spectrograph to provide individual spectra of each 0.15 arcsecond wide slice covering the 0.6 – 2.0 micron spectral range over a 3.00 x 3.15 arcsec field. The spectral resolution is 75. The instrument provides a sharp PSF, precision photometry, and stable observations for implementing the WFIRST science.

The coronagraph instrument includes an imaging mode, an integral field spectrograph, and a low order wavefront sensor to perform exoplanet detection and characterization. The coronagraph covers a spectral range of 400 - 1000 nm, providing a contrast of $10^{-9}$ with an inner working angle of $3\lambda/D$ at 400 nm. The field of view is 2.5 arcsec and the pixel scale is 0.017 arcsec. The detector is a 1k x1k EMCCD. The coronagraph includes an Integral Field Spectrograph for low-background slit spectroscopy. It operates in the 600 - 1000 nm range and has a spectral resolution of 70.

## 3. WFIRST Observations

The WFIRST observing program will be determined in detail in the year before launch to give the best science return for the dark energy, exoplanets and highest-ranked GO programs. We present here an example of the observation strategy. The program has four main components, to be carried out over six years. The microlensing survey will monitor 2.81 $deg^2$ of the Galactic bulge, with 15-minute cadence, over six 72-day seasons, detecting thousands of exoplanets via the perturbations they produce on microlensing light curves. The high latitude survey (HLS) will carry out imaging and spectroscopy of a 2400 $deg^2$ area over 1.9 years of observing time, providing weak lensing shape measurements of 400 million galaxies and emission line redshifts of more than 20 million galaxies at redshifts 1 < z < 3. The supernova survey will use 0.5 years of observing time over a 2-year period, with three tiers of imaging to discover supernovae and measure their light curve, IFU spectroscopy will be performed of more than 2700 Type Ia SNe at 0.2 < z <1.7 to measure redshifts, spectral diagnostics, and well calibrated fluxes in synthetic filter bands matched across redshift. The coronagraph direct imaging of exoplanet will take place over 1 year. The time will be divided between imaging and spectroscopy of known exoplanets from radial velocity and other programs and exoplanet searches of nearby stars. A Guest Observer program is allocated 1.5 years of observing time, spread across the 6-year mission.

Figure 1 shows the imaging depth achieved by the WFIRST HLS, in comparison to LSST (after 10 years of operation) and Euclid. In an AB-magnitude sense, WFIRST imaging is well matched to the i-band depth of LSST. The Euclid IR imaging is ~2.5 magnitudes shallower than WFIRST (and under sampled by its 0.3 arcsec near-IR pixels). The Euclid weak lensing survey uses a wide optical filter and reaches an AB-magnitude depth similar to what LSST achieves in each of its g, r, and i filters. The cumulative imaging depth in the WFIRST supernova survey fields, which will be observed many times over a 2-year interval, reaches 0.5 – 2.5 magnitudes fainter than the HLS.

Figure 2 shows the emission line sensitivity of the HLS spectroscopic survey. For moderately extended sources, the 7σ detection threshold is ≈ $1.0–1.5 \times 10^{-16}$ erg s$^{-1}$ cm$^{-2}$ depending on wavelength. The primary target is Hα at 1.05<z<2, but it is also possible to extend to higher redshift, 1.7 < z < 2.9, using galaxies with strong [OIII] emission. Forecasts based on recent estimates of the Hα and [OIII] galaxy luminosity functions predict ~ 20 million Hα galaxies, and ~ 2 million [OIII] galaxies at z > 2. At z = 1.5, the predicted comoving space density of the galaxy redshift survey is $1.5 \times 10^{-3}$ Mpc$^{-3}$. This is 16 times higher than the predicted space density for Euclid (from the same luminosity function), providing better sampling of structure, over a smaller area of sky, for studies of baryon acoustic oscillations, redshift-space distortions, and higher order galaxy clustering. The "narrow/deep" strategy of WFIRST nicely complements the "shallow/wide" strategy of Euclid.

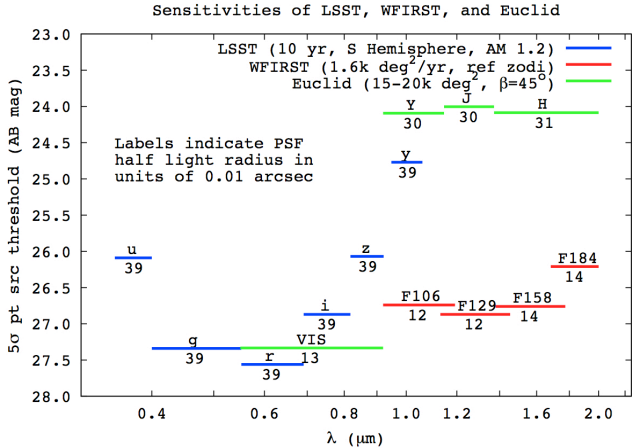
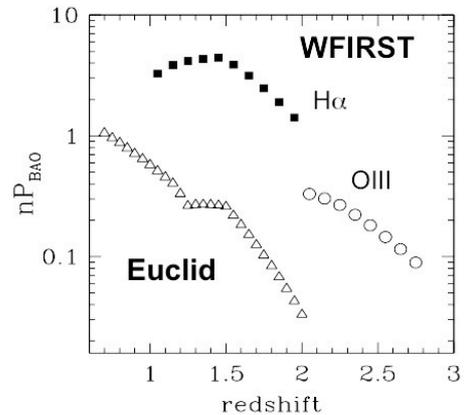

**Figure 1.** Imaging sensitivity of WFIRST compared to LSST and Euclid [1].

**Figure 2.** Spectroscopic sensitivity of WFIRST compared to Euclid [2].

## 4. WFIRST Science

*4.1. Dark Energy*
Figure 3 shows the various techniques that WFIRST will use to study dark energy. Observations of Type Ia SNe will determine "standard candle" distances out to z = 1.7, calibrated against a (ground-based) sample observed in the local Hubble flow. The aggregate precision of these measurements is 0.20% at z < 1 (error-weighted <z> = 0.50) and 0.34% at z > 1 (<z> = 1.32). The baryon acoustic oscillation (BAO) feature in galaxy clustering provides a "standard ruler" for distance measurement, calibrated in absolute units, independent of $H_0$. The galaxy redshift survey (GRS) enables measurements of the angular diameter distance $D_A(z)$ and the expansion rate $H(z)$ using H$\alpha$ emission line galaxies at z = 1 - 2 and [OIII] emission line galaxies at z = 2 – 3, with aggregate precision ranging from 0.40% to 1.8%.

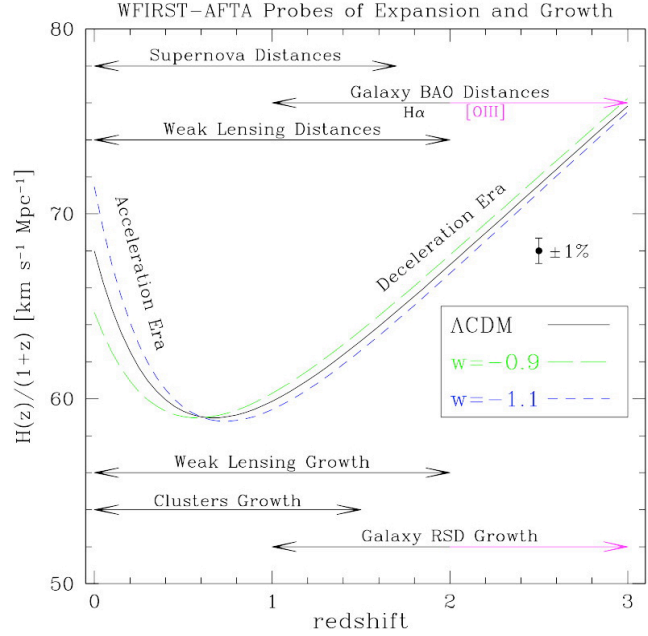

**Figure 3.** Dark energy techniques used by WFIRST [2].

The imaging survey will enable measurements of dark matter clustering via cosmic shear and via the abundance of galaxy clusters with mean mass profiles calibrated by weak lensing; we expect 40,000 M ≥ $10^{14}M_{Sun}$ clusters in the 2400 $deg^2$ area of the high- latitude survey. These data constrain the amplitude of matter fluctuations at 0 < z < 2 and provide additional leverage on the redshift-distance relation. The expected aggregate precision on the fluctuation amplitude as an isolated parameter change is ≈ 0.15% at z < 1 and 0.3- 0.5% at z > 1. Redshift-space distortions in the GRS provide an entirely independent approach to measuring the growth of structure, with aggregate precision ≈ 1% at z = 1-2.

*4.2. Microlensing*
The WFIRST microlensing survey will continuously monitor a total of 2.81 $deg^2$ in the Galactic bulge for six 72-day campaigns, with a cadence of 15 minutes in a wide filter (0.927 – 2.00 μm) for planet discovery and 12 hours in z-band for characterization of source and lens stars. This survey will detect tens of thousands of stellar microlensing events and, in cases where the lensing star has an orbiting planet, can (when the alignment is favorable) produce a distinctive perturbation to the light curve. For a large fraction of these events, WFIRST will be able to completely characterize the lensing system, including the host star and planet mass, the planet orbital separation in physical units, and the distance to the system. It will do this through a combination of subtle deviations from the simplest light curve shapes, including finite-source effects, "parallax" caused by motion of the earth and the observatory's geosynchronous orbit, and source astrometric motion induced by the host lens, together with the isolation and direct measurement of the light from luminous host lenses afforded by the high angular resolution.

Figure 4 shows the parameter space for the WFIRST microlensing program compared to Kepler. Forecasts suggest that WFIRST will detect about 3000 planets overall, including 1000 "super-earths" (roughly 10 times the mass of earth), 300 earth-mass planets, and 40 Mars-mass planets. These detections would enable measurement of the mass function of cold exoplanets to better than ~10% per decade in mass for masses > 0.3 $M_{Earth}$ and an estimate of the frequency of Mars-mass embryos accurate to ~15%. If there is one free-floating earth-mass planet per star in the Galaxy, then WFIRST will detect about 40 of them, and much larger numbers of more massive free-floating planets.

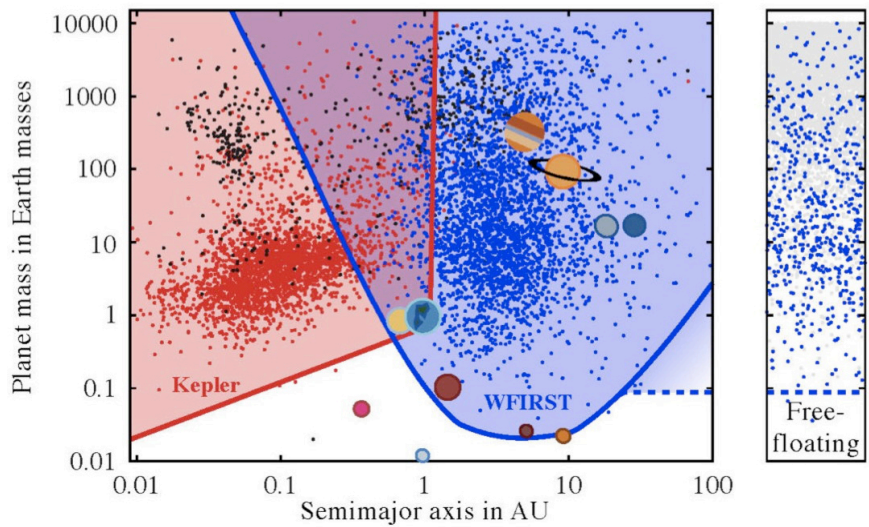

**Figure 4.** Parameter ranges for exoplanet detection for WFIRST and Kepler [2].

*4.3. Exoplanet Direct Imaging*
With a contrast level of $10^{-9}$ and an inner working angle of less than 0.2 arcsec, the WFIRST coronagraph will observe known nearby exoplanets and perform a survey of the nearest stars to search for exoplanets. These observations will directly image over a dozen known radial velocity planets and will discover an additional dozen previously unknown ice and gas giants as shown in Figure 5. The majority of these planets will also be characterized using R~70 spectra in the wavelength range 600-1000 nm via the IFS. These spectra will allow the detection of features expected due to methane, water, and alkali metals, reveal the signature of Rayleigh scattering, and easily distinguish between different classes of planets (i.e., Neptunes versus Jupiters), and planets with different metallicities. This survey will also be sensitive to debris disks with a few times the solar system's level of dust in the habitable zones and asteroid belts of nearby (~10 pc) Sun-like stars.

The high sensitivity and spatial resolution (0.05 arcsec is 0.5 AU at 10 pc) of WFIRST images will map the large-scale structure of these disks, revealing asymmetries, asteroid belts, and gaps due to unseen planets. WFIRST will make the most sensitive measurements yet of the amount of dust in or near the habitable zones of nearby stars. This is important for assessing the difficulty of imaging Earth-like planets with future missions as well as for understanding nearby planetary systems. Finally, spectrophotometry of these disks at the full range of available wavelengths from 400-1000 nm provides constraints on dust grain size.

*4.4. Community Science*
The combination of a 0.28 $deg^2$ field with the near-IR sensitivity and angular resolution afforded by a 2.4 m telescope gives WFIRST extraordinary discovery potential for general astrophysics. In the 6-year prime mission, 1.5 years would be allocated to GO programs. If WFIRST enters an extended mission phase, observing time would be competed, with the balance between large and small programs to be decided by the Time Allocation Committee. This will afford a large number of unique, high impact projects from characterizing nearby brown dwarfs to probing the epoch of reionization.

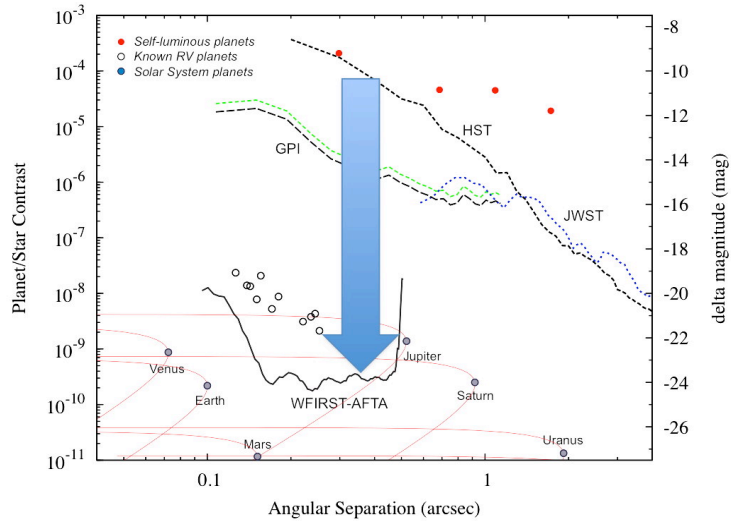

**Figure 5.** Exoplanet detection sensitivity of the WFIRST coronagraph compared with other instruments [2].

The WFIRST HLS will probe the low surface brightness structure around many thousands of nearby galaxies, and GO observations can extend these investigations to greater depth and to systematic coverage of different environments, from the centers of voids to the core of the Virgo cluster. For galaxies within a few Mpc, WFIRST will resolve individual red giant stars as well as imaging the low surface brightness background of main sequence stars. These investigations will show how recent accretion and minor mergers affect the growth of galaxies across a wide spectrum of mass, morphology, and environment. Regions of star formation, for example, the Taurus association, stretch across many degrees of the sky. The near-IR sensitivity of WFIRST will make possible a full census of stars in these systems to a depth far beyond what has been possible with ground-based and space-based telescopes. There are many other examples of community science, including white papers written by dozens of astronomers, in the 2013 SDT report [1].

## 5. WFIRST and Gravitational Wave Electromagnetic Counterparts

The next ten years will see major steps forward in the search and detection of gravitational waves. LIGO and Virgo will be in their advanced detector era with sensitivity by 2020 to detect NS binary mergers to 200 Mpc [3]. The detection rate will be 10's of events per year, with large uncertainties. In the next few years, the pulsar timing array [4] will be observing a sufficient number of millisecond pulsars with accurate enough timing observations to begin detecting binary supermassive black holes in very low frequency gravitational waves. The LISA Pathfinder mission [5] will launch in 2015 and demonstrate the drag-free control and laser interferometry technologies needed for LISA. For all of these current and future instruments, there is a great interest in detecting electromagnetic counterparts of their gravitational wave sources in order to determine distances and astrophysical properties. WFIRST can play a role in the counterpart observations.

For LIGO-Virgo, the most likely early gravitational wave detections will be NS-NS or NS-BH mergers within a range of 50 – 200 Mpc. They will be accompanied by bright gamma-ray bursts and afterglow if the jet axis is aimed at the observer. The chance of such alignment is about 1%. In the more common off-axis case, the event will be accompanied by a faint afterglow or possibly faint kilonova emission from radionuclides produced in the dense neutron-rich material surrounding the merger site. The kilonova emission is predicted to peak in the NIR, where WFIRST is the most sensitive. This is due to the opacity from the heavy elements formed in the ejecta which would be much more neutron rich than the ejecta in a type Ia supernova, where the opacity is dominated by Fe. [6]

The sky locations will not be accurate for early LIGO-Virgo events with error box sizes of 100's of deg$^2$ [3]. This will improve to a few deg$^2$ in the WFIRST time frame when LIGO-India is on-line. There will be several ways in which WFIRST can participate in follow-up observations as listed here:
1. Tile the error box with the WFI to search for a new source. With a 0.28 deg$^2$ field of view, 10's of pointing will be required. The afterglows could be in the one to two orders of magnitude brighter than the WFIRST J=27 AB sensitivity for days so detections are possible.
2. Perform pointed observations of galaxies within the error box. Since the mergers will occur in or near galaxies, WFIRST can point at the few brightest galaxies in the error box to search for emission [7]. This can will also be possible with the JWST.
3. Perform deep follow-up studies of candidate electromagnetic sources. WFIRST and JWST can observe candidate sources found by other telescopes and perform deep imaging and spectroscopic observation in the NIR.

For LISA, the most likely gravitational wave detections will be binary SMBH mergers at cosmic distances. Bare BH mergers have no electromagnetic radiation, but gas around the SMBHs will be stirred up and accrete forming quasar-like activity on years to decades time scales. Also, stars around the SMBHs will have their orbits perturbed and create tidal disruption events on years to decades time scales.

The sky location accuracy for LISA will like be a few deg$^2$ [8]. The primary way that WFIRST or a future similar telescope can make observations of LISA sources will be to perform monitoring observations of the fields to search for fading or tidal disruption flares. The error circle will either need to be tiled or candidate galaxies selected for pointed observations.

## 6. Conclusions

WFIRST will be a powerful mission for NIR surveys and exoplanets. It will perform Hubble quality and depth imaging and spectroscopy over 1000's of square degrees. It will advance our understanding of the expansion history of the universe and growth of structure driven by dark energy using several techniques. It will make a large step forward in exoplanet research by performing the first space microlensing census observing the dense star population in the galactic bulge and the first high-contrast coronagraphic direct-imaging observations of nearby exoplanets and debris disks. It will have a vigorous and well-fund guest observer program for community astrophysics. With its NIR coverage, depth and wide field of view, WFIRST will be a powerful tool for follow-up of LIGO-Virgo and LISA gravitational events.